\documentclass[twocolumn,prb,showpacs,preprintnumbers,amsmath,amssymb]{revtex4}

\usepackage{graphicx}
\usepackage{dcolumn}
\usepackage{bm}
\usepackage{multirow}

\begin{document}
\preprint{APS/123-QED}
\title{Role of Auger Recombination in Plasmon Controlled Photoluminescence Kinetics in Metal-Semiconductor Hybrid Nanostructures}
\author{Sabina Gurung$^{ab}$}
\email{sabusg12@gmail.com}
\author{Asha Singh$^{a}$}%
\author{J. Jayabalan$^{ab}$}
\affiliation{$^{a}$ Nano Science Laboratory, Materials Science Section, 
	Raja Ramanna Centre for Advanced Technology, Indore, India - 452013. \\
		$^{b}$ Homi Bhabha National Institute, Training School Complex, Anushakti Nagar, Mumbai, India - 400094.}
\date{\today}

\begin{abstract}
Spectroscopic studies of semiconductor quantum dots (SQDs) addressing the problem of non-radiative carrier losses is vital for the improvement in the efficiency of various light-emitting devices. Various designs of SQDs emitter like doping, forming core-shell and alloying has been attempted to suppress non-radiative recombination. In this article, we show that forming a hybrid with metal nanoparticles (MNP) having localized surface plasmon resonance overlapped with the emission spectrum of SQD, the non-radiative carrier loss via Auger recombination can be mitigated. Using steady-state and time-resolved photoluminescence, it has been shown that when such hybrid is selectively excited well above the bandgap without exciting plasmon, the contribution to fast decay time reduces along with an increase in contributions to longer decay times. A completely reverse kinetics is observed when exciton and plasmon are simultaneously excited. Such control of photoluminescence kinetics by placing MNP near SQD opens up a new method for designing hybrid materials that are well suited for light-emitting devices.
\end{abstract}

\maketitle

\section{Introduction}
Over the past few decades, semiconductor quantum dots (SQDs) have become a valuable component in various optoelectronics devices. Their size dependent properties like absorption or photoluminescence (PL) spectrum allows excellent control over their properties for light emission and light-harvesting applications \cite{Luminescent_CdTe_QD_2003, Size_selective_PL_CdTeQD_2003, Guo-ReactionConditions-TGA-CdTeNP-JPCB-2005}. Several methods are being attempted to enhance and control the PL emission kinetics of SQDs for the development of efficient optoelectronic devices\cite{Tailoring_Properties_crystanality_Nature_2007, photophysical_properties_Au_CdTe_varying_size_shape_2012}. Recently, decorating the SQD with metal nanoparticle (MNP) or vise-versa to form hybrid nanostructures has gained considerable attention due to the flexibility it offers in controlling parameters like size, shape and spatial distribution\cite{Enhancement_quenching_metal_sc_hybrid_Viste_2010, metal_sc_plasmon_enhanced_Jiang_Adv_matter_2014, metal_Sc_hybrid_selforganization_exciton_plsmon_Strelow_2016}.
Independently, MNP shows an unusual resonance known as the localized surface plasmon resonance (LSPR), lies mostly in the visible regime. When excited at the LSPR, the free electrons inside the MNP oscillates in-phase with the applied field creating a high local field around it\cite{Gustav_Mie_LSPR_1908}. When a SQD is placed near to such MNP, the plasmon oscillations strongly modifies the response of SQD to the applied field\cite{Gustav_Mie_LSPR_1908,LSPR_review_Analytica_chimica_2011, Local_field_spontaneous_rate_CdTe_CdSe_2004}. The field enhancement due to LSPR can lead to an enhancement in the absorption, emission cross-section and Raman scattering from the nearby SQD\cite{LSPR_Bisosensing_Ruemmele_ACS_2010,Nanoplasmonics_enhancement_Nanotechnology_2006,Plasmon_solar_energy_Scott_2013,EMAM2018287}. It is also possible that an electron-hole recombination can occur in an excited SQD, non-radiatively by exciting a plasmon in the MNP causing an reduction in PL emission\cite{Ag_cdte_selforganized_electrostatic_interaction_Wang_Spect_Acta_2005,Ultrafasr_charge_Transfer_Samanta_2016,Sabina-JAP-2018}. On the other hand, a plasmon excited in the MNP can decay by exciting a electron-hole pair in SQD and thereby contribute to an increase in PL\cite{Controlling_PIRET_Metal_TiO2_JPCC_2015}. Such energy exchange will strongly depend on several factors like shape, size, properties of the individual elements and its distribution\cite{Energy_transfer_Qd_Au_2004, Size-Dependent_energy_transfer_CdSe_ZnSQD_Au_2011, KANEMITSU2011510, Surface_energy_transfer_JPCC_2018,LSPR_spectraloverlap_energy_transfer_dye_Au_2010}. Consequently, a suitably designed MNP-SQD hybrid nanostructures(HNSs) can display extraordinary optical properties with multiple functionalities that are derived from the synergistic interactions between MNPs and SQDs.

Controlled growth of individual nano-components and self-organized growth of HNS of specific structure has provided an opportunity to study the effect of structure on PL kinetics\cite{Sabina-JAP-2018, Ag_cdte_selforganized_electrostatic_interaction_Wang_Spect_Acta_2005}. Most of these studies in HNSs have been focused on exploring distance or size-dependent plasmon-exciton coupling between MNP and SQD\cite{Enhancement_quenching_metal_sc_hybrid_Viste_2010, metal_sc_plasmon_enhanced_Jiang_Adv_matter_2014, metal_Sc_hybrid_selforganization_exciton_plsmon_Strelow_2016}. There are very few reports on the effect of excitation energy on the relaxation process in SQDs in the presence of MNP. The observed modification in PL kinetics in the presence of MNP has been attributed to different coupling mechanisms like charge transfer, energy transfer, increased multiphoton absorption due to LSPR and so forth\cite{Charge_transfer_Au_CdSe_Gao_2012, LSPR_spectraloverlap_energy_transfer_dye_Au_2010, Plasmon_effect_multiexciton_QD_Zhao_2016, Multiphoton_enhancement_CdSe_Au_film_Moyer_2013}. However impact of excitation energy on the relaxation process in SQD in the presence of MNP still remains unclear. 

In this article, we report a study on the effect of excitation energy on the PL kinetics of HNSs designed specifically with and without overlap between PL emission and LSPR absorption. Both PL spectra and time-resolved PL were studied by exciting at different photon energies to understand the changes in PL kinetics. It was observed that a selective excitation of exciton-plasmon could drastically alter the PL kinetics. Such tuning of PL kinetic in HNS find its potential application in designing light-harvesting devices that require enhanced radiative recombination rate and prolonged lifetime as well as optoelectronic devices, which require shorter lifetime \cite{LED_MRSbulletin_2013}.

\section{Results and Discussion}
\begin{figure}[h]
	\includegraphics[width=0.9\columnwidth]{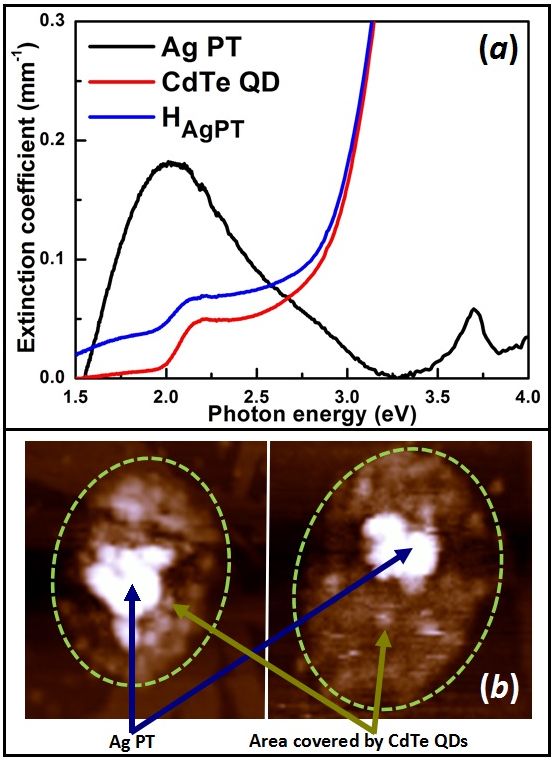}
	\caption{({\it a}) Extinction spectrum of the colloidal solutions of Ag PT, CdTe QD and H$_{AgPT}$. ({\it b}) AFM topography image of the H$_{AgPT}$ hybrid colloid. \label{FigExtCoeff}}
\end{figure}

Individual colloidal samples of CdTe quantum dots (QDs) and silver nanoplatelets (PT) dispersed in water were prepared by wet chemical techniques\cite{Silver_sphere_Science_2001, Abhijit-CdTe-Pre-PL-JPCC-2008}. The procedure followed for the preparation of these individual colloids have already been reported in our earlier articles\cite{Sabina-JAP-2018,Counting_electron_Jayabalan_2019}. A short description of these preparation procedures are also given in the supporting information (SI). In the final colloidal solutions, the CdTe QDs are capped by thiolglycolic acid (TGA), while the Ag PT are capped by trisodium citrate and polyvinylpyrrolidone (PVP). The volume fraction of CdTe QD and Ag PT in their corresponding colloidal solution is of the order of 10$^{-6}$ and 10$^{-7}$ respectively. At these volume fractions, the individual particles are well separated and the particle-particle interactions is negligible. Figure\ref{FigExtCoeff}({\it a}) shows the extinction coefficient of the colloidal solutions of Ag PT and CdTe QD. In case of CdTe QD, the lowest energy excitonic peak, the 1s-1s transition, lies at 2.21 eV\cite{Pengs_cdte_quantum_dots_size_2003,UCPL_CdSe_2005}. The size of the CdTe QDs estimated using Peng's formula from its excitonic peak position is 3.4 nm\cite{Pengs_cdte_quantum_dots_size_2003}. The topography measurement of CdTe QDs on a mica substrate shows well separated individual particles and confirms this estimated size (Fig.S1({\it a})). TEM measurement of Ag PT shows particles of an average diameter and thickness 34 nm and 5.2 nm respectively (Fig.S1({\it b})). The extinction spectra of the Ag PT colloid shows an in-plane dipole LSPR peak around 2.05 eV and an out-of-plane quadruple peak at 3.69 eV (Fig.\ref{FigExtCoeff}({\it a}))\cite{JJ-APL-Edgesmoothening}. 

The colloidal solution of the hybrid nanostructure is prepared by mixing 5 ml of the as-prepared CdTe QD colloid and 1 ml of Ag PT colloid. Based on our previous study, it is expected that at this concentration in the H$_{AgPT}$ hybrid colloid, each of the Ag PT is completely surrounded by CdTe QDs such that no further CdTe QD attachment is possible\cite{Sabina-JAP-2018}. Figure\ref{FigExtCoeff}({\it b}) shows AFM topography image of a drop of H$_{AgPT}$ dried on a mica substrate. Unlike the image of bare CdTe sample, several aggregated particles can be seen  in the topographical image of H$_{AgPT}$ . In each of these aggregates, a large particle of dimension comparable to that of Ag PT is found to be surrounded by several small particles of dimension close to that of CdTe QDs. The capping agent of CdTe QD, the TGA, has a higher affinity towards metal surface compared to that of citrate\cite{Ligand_exciton_photovoltaic_PbS_Jin_PCCP_2017, TGA_CdTe_Europium_surface_coordinated_emission_Gallagher_inorganic_chemistry_2013}. Since the TGA molecule is already attached to a CdTe QD, it holds both MNP and SQD together when it gets attached to the metal surface. It is known that the properties of hybrid strongly depends on the separation
between the MNP and SQD\cite{Govorov_NanoLett_2006,Precise_control_JAP_2013,Viste_ACSNano_2010}. In the literature, it isreported that the length of the TGA is about 0.5 nm; thus, the separation between the CdTe QD and Ag PT is expected to be nearly 0.5 nm\cite{Ligand_exciton_photovoltaic_PbS_Jin_PCCP_2017, TGA_CdTe_Europium_surface_coordinated_emission_Gallagher_inorganic_chemistry_2013}. Since the surface area of Ag PT is much larger than that of CdTe QD, several CdTe QDs could get attached to a single Ag PT. The aggregated structures seen in the AFM images are due to the single Ag PT surrounded by several small CdTe QDs. The extinction spectra of colloidal solution of hybrid sample H$_{AgPT}$ is also shown in Fig.\ref{FigExtCoeff}({\it a}) and is clearly different from that of both the individual CdTe QD and Ag PT colloids. An estimated
extinction spectrum of a sample which contains a non-interacting mixture of Ag PTs and CdTe QDs does not match the experimentally measured spectrum of H$_{AgPT}$. This dissimilarity indicates that there is an interaction between the Ag PT and CdTe QD which results in modifying their individual properties. The extinction spectra of the H$_{AgPT}$ colloid looks more like that of CdTe QD but with two modified features, (i) the bleaching of the exciton peak at around 2.21 eV and (ii) band tailing at the lower energy side below 2 eV. Bleaching of exciton peak in the presence of MNP has been attributed to exciton dissociation in the presence of high electric fields \cite{XIA2008166, Enhanced_PL_CdTe_Ag}. On the other hand the band tailing below $2$ eV is also observed in several other metal-semiconductor HNS and has been attributed to the formation of additional surface defect states\cite{Nahar_ACSNano_2015}. The increase in band tailing in the H$_{AgPT}$ sample suggests an increase in surface defect states during the process of hybrid formation by self-organization\cite{Luminescent_CdTe_QD_2003}. 

\begin{figure}[h]
	\centering
	\includegraphics[width=0.9\columnwidth]{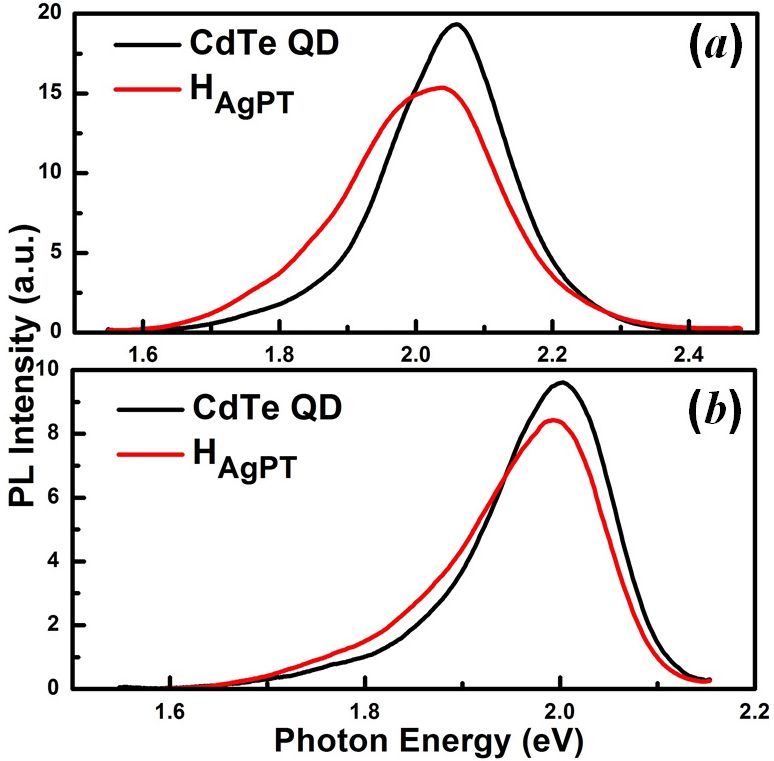}
	\caption{Photoluminescence spectra of CdTe QD and H$_{AgPT}$ colloids when excited at ({\it a}) 3.06 eV and ({\it b}) 2.25 eV. \label{PL}}
\end{figure}%

The time-resolved PL measurements were carried out by exciting the sample using the output of an optical parametric amplifier (OPA). The OPA was pumped by a 1 kHz Ti:Sapphire amplifier system with a pulse duration of 35 fs at 800 nm wavelength. The output of the OPA, either tuned to 405 nm (3.06 eV) or 550 nm (2.25 eV), was used for the selective excitation of the samples. The pulse width of the OPA output was nearly $\sim$ 80 fs. The intensity of the laser beam at the sample position was kept nearly 150 MW/cm$^{-2}$. All the measurements were carried out at room temperature. Before looking at the PL kinetics, let us first examine the effect of the presence of Ag PT near CdTe QD on its PL spectrum. Figure\ref{PL} ({\it a}) and ({\it b}) shows the measured PL spectra of the bare CdTe QD and H$_{AgPT}$ colloids when excited at 3.06 eV (405 nm) and 2.25 eV (550 nm) respectively. The PL spectrum is nearly Gaussian with a longer tail on the lower energy side. When compared to the PL emission from CdTe QD, the hybrid sample H$_{AgPT}$ shows three well distinct changes: (i) the strength of the PL at its peak emission got quenched at both these excitation energies, (ii) there is an enhancement in the PL emission at the lower energy side and (iii) the total integrated area under the PL spectrum remains nearly same compared to that of corresponding CdTe QD colloid i.e.102\% and 96\% for 3.06 eV and 2.25 eV excitation respectively. 

It is well known that the presence of MNP can lead to quenching or enhancement in the PL emission from a SQD placed nearby. Several factors like nature of the individual materials, shape of the final nanoaggregate, distance between the metal and semiconductor, the nature of junction between them and linking medium are known to play a role in enhancing or quenching the PL emission\cite{Enhancement_quenching_metal_sc_hybrid_Viste_2010,metal_sc_plasmon_enhanced_Jiang_Adv_matter_2014,metal_Sc_hybrid_selforganization_exciton_plsmon_Strelow_2016}. Presence of MNP is also known to introduce additional defects states in CdTe QD due to ligand exchange while forming the hybrid nanostructures\cite{Ligand_exciton_photovoltaic_PbS_Jin_PCCP_2017, TGA_CdTe_Europium_surface_coordinated_emission_Gallagher_inorganic_chemistry_2013}. In the present case, the quenching of PL strength at the peak along with an enhancement in the PL at the lower energy side clearly indicates that part of the charge carriers from the band edge are relaxing to the newly formed defect states from which they further decay radiatively. The fact that the integrated area under the PL spectrum of H$_{AgPT}$ remains same as that of bare CdTe QD colloids indicates that nearly all carriers relaxed to the defect states are able to relax radiatively.

\begin{figure}[h]
	\centering
	\includegraphics[width=0.9\columnwidth]{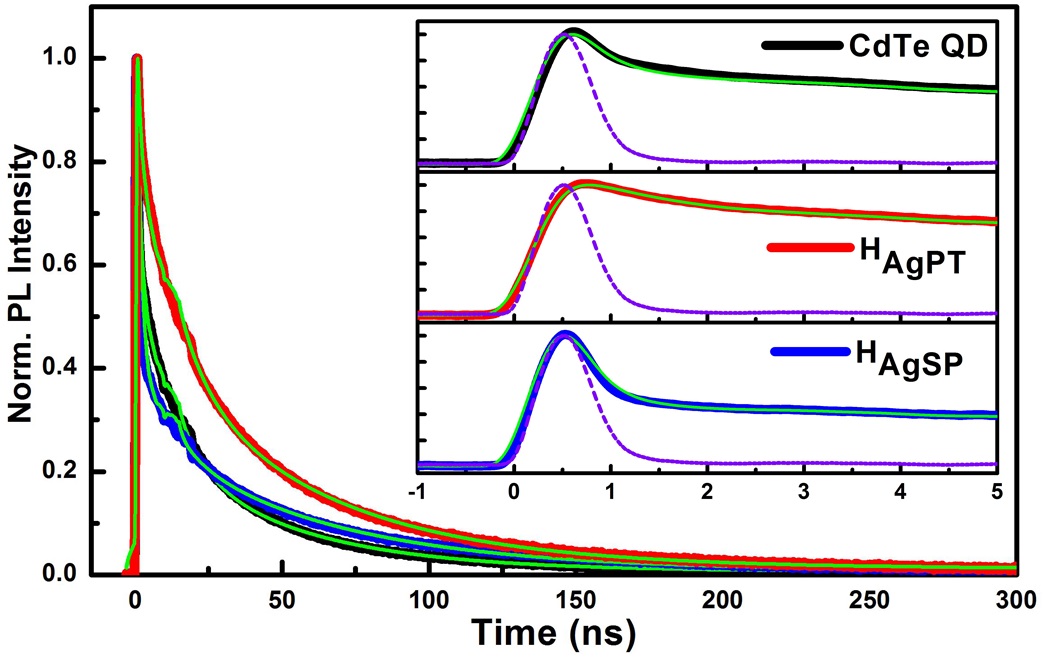}
	\caption{Time-resolved PL decay kinetics of CdTe QD, H$_{AgPT}$ and H$_{AgSP}$ colloids when excited at 3.06 eV. Inset shows the time-resolved PL decay kinetics zoomed in time. \label{405nm_lifetime}}
\end{figure}%

\begin{figure}[h]
	\centering
	\includegraphics[width=0.9\columnwidth]{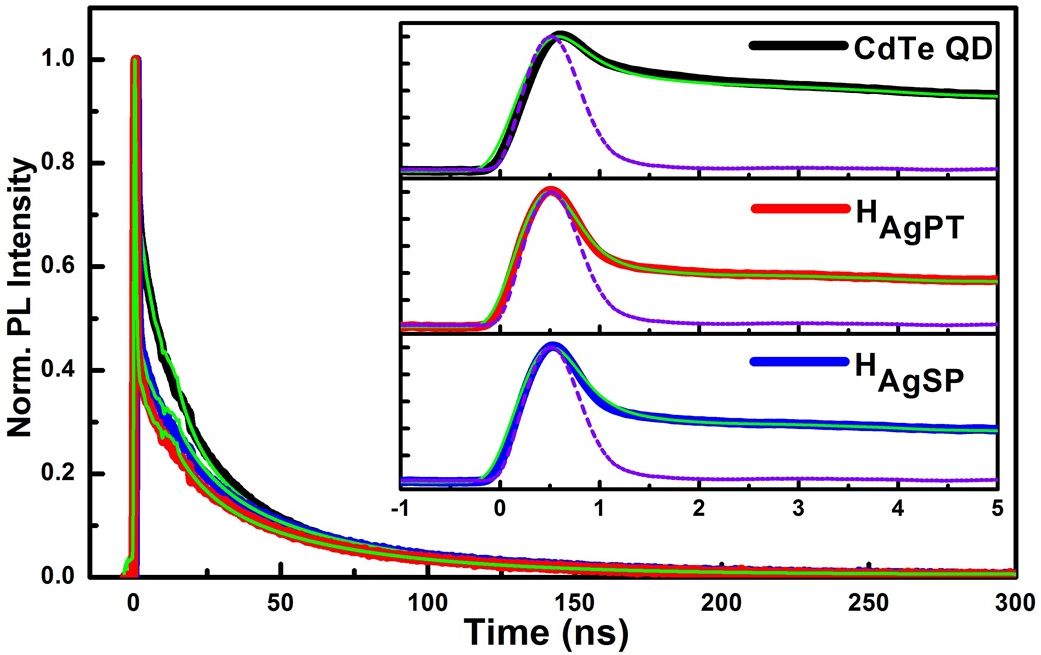}
	\caption{Time-resolved PL decay kinetics of CdTe QD, H$_{AgPT}$ and H$_{AgSP}$ colloids when excited at 2.25 eV. Inset shows the time-resolved PL decay kinetics zoomed in time. \label{550nm_lifetime}}
\end{figure}

The time-resolved PL measurement was carried out using a fast photodetector-oscilloscope system at room temperature. The photodetector had a rise time of 300 ps. The output of the detector was measured by an oscilloscope having a bandwidth of 1 GHz and a sampling rate of 20 GS/s. The decay time estimated for the later part of the instrument response function (IRF) is 350 ps. Figure\ref{405nm_lifetime} and \ref{550nm_lifetime} shows that the temporal evolution of the PL of CdTe QD colloid when excited at 3.06 eV and 2.25 eV respectively. The measured IRF are also shown in the inset figures. With the arrival of the excitation pulse, the PL emission starts and over the next few tens of ns the PL decays at different rates. Depending on the number of processes that control PL emission, the II-VI semiconductor QD colloid shows a bi- or tri-exponential decay of the PL \cite{TGA_CdTe_2007, InfluenceofAcid_PL_TGA_CdTe_Mandal_2008,Surface_related_emission_CdSe_Wang_2003,Effect_Chloride_Passivation_CdTe_Binks_2015}. We find that the temporal evolution of PL measured for CdTe QD colloid when excited at both the photon energies fits well to a tri-exponential decay function convoluted ($\otimes$) with the experimental IRF ($I_{IRF}$), given by
\begin{eqnarray}
F(t) = \left(I_{IRF} \otimes \sum_{i = 1,2,3}{A_{i} e^{\frac{-t}{\tau_{i}}} }\right)
\label{Eq:TriExpNSPL}
\end{eqnarray}
where A$_{i}$ is the i$^{th}$ signal amplitude corresponding to the time constant (decay time) $\tau_{i}$. The best fit parameters obtained by fitting Eq.\ref{Eq:TriExpNSPL} to the experimental CdTe QD PL is summarized in Table-\ref{Table1-nsPLdecay}. 

\begin{table*}[]
\caption{Best fit parameters obtained by fitting Eq.\ref{Eq:TriExpNSPL} to the experimental time-resolved PL of colloidal solutions of CdTe QDs, H$_{AgPT}$ and H$_{AgSP}$.} \label{Table1-nsPLdecay} 
\def\arraystretch{1.0}
\setlength{\tabcolsep}{1.8em}
\begin{tabular}{|c|c|c|c|c|c|c|c|}
\hline
Sample          & \begin{tabular}[c]{@{}c@{}}Excitation\\ Energy (eV)\end{tabular} & \begin{tabular}[c]{@{}c@{}}$A_{1}$ \\ \%\end{tabular} & \begin{tabular}[c]{@{}c@{}}$\tau_{1}$\\ (ns)\end{tabular} & \begin{tabular}[c]{@{}c@{}}$A_{2}$ \\ \%\end{tabular} & \begin{tabular}[c]{@{}c@{}}$\tau_{2}$\\ (ns)\end{tabular} & \begin{tabular}[c]{@{}c@{}}$A_{3}$\\ \%\end{tabular} & \begin{tabular}[c]{@{}c@{}}$\tau_{3}$\\ (ns)\end{tabular} \\ \hline
\multirow{2}{*}{CdTe QD} & 3.06                               & 77                        & 0.23$\pm$0.02                       & 14                        & 7$\pm$ 0.2                         & 9                        & 65$\pm$0.2                        \\ \cline{2-8} 
             & 2.25                               & 72                        & 0.23$\pm$0.01                       & 17                        & 7$\pm$0.1                         & 11                       & 65$\pm$0.3                        \\ \hline
\multirow{2}{*}{H$_{AgPT}$}  & 3.06                               & 34                        & 0.7$\pm$0.02                        & 38                        & 8.7$\pm$0.2                        & 28                       & 200$\pm$0.8                        \\ \cline{2-8} 
             & 2.25                               & 91                        & 0.15$\pm$0.001                       & 7                        & 6.5$\pm$0.4                        & 4                        & 120$\pm$1.3                        \\ \hline
\multirow{2}{*}{H$_{AgSP}$}  & 3.06                               & 87                        & 0.21$\pm$0.002                       & 7                        & 6.5$\pm$0.2                       & 6                        & 200$\pm$0.6                     \\ \cline{2-8} 
             & 2.25                               & 87                        & 0.19$\pm$0.003                       & 7                        & 6$\pm$0.2                         & 6                        & 150$\pm$0.5                        \\ \hline
 
\end{tabular}
\end{table*}

At 3.06 eV excitation, a strong portion of the PL, 77\%, decays with a time constant of 0.23 ns, 14\% decays with a time constant of 7 ns, and about 9\% of the PL decays with a much longer lifetime of $\sim$ 65 ns. The best-fit parameters (decay times and amplitudes) for the bare CdTe QDs at excitation energy, 2.25 eV, also remain almost similar to that of 3.06 eV excitation. This similarity indicates that the PL relaxation processes does not depend on the excitation photon energy for CdTe QD. A similar observation was also reported by Dey {\it et al.} and can be explained by the anti-bunching nature of the photons emitted from single QD \cite{Antibunched_QD_Zhao_2018}. Figure\ref{405nm_lifetime} also shows the time dependence of the PL measured for the hybrid sample, H$_{AgPT}$, when excited at 3.06 eV. The PL measured for this case also fits well to the tri-exponential decay function given by Eq.\ref{Eq:TriExpNSPL}. The best-fit parameters estimated for this case are also summarized in Table-\ref{Table1-nsPLdecay}. In case of H$_{AgPT}$, the three different time constant, $\tau_{1}$, $\tau_{2}$ and $\tau_{3}$ turns out to be 0.70 ns, 8.7 ns and 200 ns respectively. The shortest and longest decay times of H$_{AgPT}$ increased by approximately three times compared to that of the bare CdTe QD, while the $\tau_{2}$ showed only a slight increase. The amplitude of each decay components, $A_i$'s, has also changed dramatically compared to that of bare CdTe QD colloid. The amplitude of the fast lifetime component ($A_{1}$) has substantially decreased while the other two amplitudes, A$_{2}$ and A$_{3}$, have increased. Now let us compare the behavior of H$_{AgPT}$ when excited at 2.25 eV with that of bare CdTe QDs. Figure\ref{550nm_lifetime} shows the temporal evolution of the PL of H$_{AgPT}$ colloid when excited at 2.25 eV (see Table-\ref{Table1-nsPLdecay} for best fit parameters). When compared to CdTe QD colloid, the $\tau_1$ of H$_{AgPT}$ becomes even shorter, 0.15 ns, with a strong increase in the corresponding amplitude $A_1$. This is completely opposite to that when excited at 3.06 eV, where it has become slower alongwith a reduction in amplitude compared to that of CdTe QD. The $\tau_{2}$ once again shows only a slight change when compared to that of CdTe QD and the $\tau_{3}$ has increased by a factor of two.

\begin{figure}[h] 
	\centerline{\includegraphics[width=1\columnwidth]{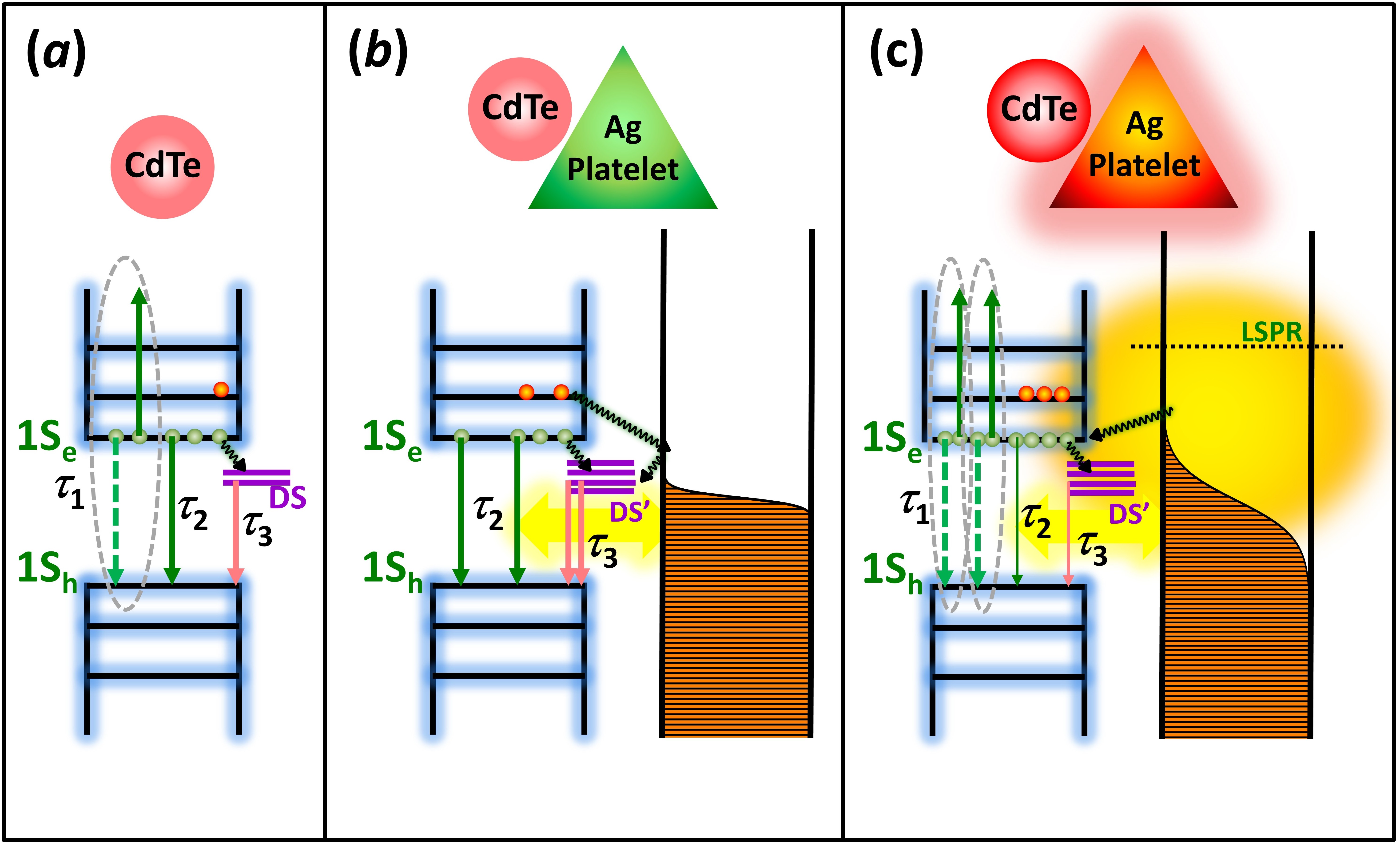}}
	\caption{Schematic illustration of PL kinetic processes in ({\it a}) CdTe QD colloid and ({\it b}) \&  ({\it c}) H$_{AgPT}$ colloid excited at 3.06 eV and 2.25 eV respectively. The dotted circle represents the Auger process, wavy and solid arrows represent non-radiative and radiative decay path respectively. DS represents defect states in the bare CdTe QD colloid and DS$^{'}$ represents the defect states in H$_{AgPT}$.  \label{Schematic}}
\end{figure}

Several groups have studied the origin of PL and the decay process in bare CdTe QDs \cite{TGA_CdTe_2007,Effect_Chloride_Passivation_CdTe_Binks_2015,InfluenceofAcid_PL_TGA_CdTe_Mandal_2008}. When the QD is illuminated by a light pulse having photon energy above the bandgap, carriers are excited in the QD such that electrons and holes are excited to the conduction and valence bands respectively. In the present case, the 1s-1s exciton transition energy is about 2.21 eV (Fig.\ref{FigExtCoeff}). When illuminated by a 3.06 eV photon, electrons and holes are excited deep into the conduction band and valence band respectively. In the next few hundreds of femtoseconds, these electrons and holes relax non-radiatively to the corresponding band edge to finally form excitons. These excitons relaxes by the emission of photons. Such radiative recombination takes place over a period of several nanoseconds in similar QDs \cite{Luminescent_CdTe_QD_2003,Size_selective_PL_CdTeQD_2003,Guo-ReactionConditions-TGA-CdTeNP-JPCB-2005}. During this radiative recombination, a part of the carriers may relax to defect states present in the sample. Such transfer of carriers from the band edge to the defect state will reduces the excitonic emission. If the carrier transfer rate to the defect state is smaller than the radiative recombination lifetime, the PL kinetics will also show the signature of this relaxation. A fast decay time in the order of 0.1 ns to 1.5 ns has been observed in CdTe QD and has been attributed to the non-radiative Auger recombination\cite{Auger_CdTe_PCCP_2013, Effect_Chloride_Passivation_CdTe_Binks_2015}. In the Auger mechanism, when there is more than one carrier excited in the QD, one of the carriers can recombine non-radiatively by transferring its energy to excite the other carrier deeper into the band or to a defect state\cite{Auger_CdS_JPCL_2011,Auger_CdTe_PCCP_2013}. We attribute the shortest relaxation time observed, $\tau_1$, in the case of CdTe QD colloid to the Auger recombination. On the other hand, in the present case of CdTe QD, we attribute the process responsible for the decay times, $\tau_{2}$ and $\tau_{3}$, to the radiative electron-hole recombination and trap-state emission respectively\cite{Effect_Chloride_Passivation_CdTe_Binks_2015,Surface_related_emission_CdSe_Wang_2003,PL_Upconversion_CdTe_Xiao_2003}. Figure\ref{Schematic}({\it a}) shows the schematic representation of different processes of PL kinetics in CdTe QD.

As mentioned earlier, when a MNP is brought close to the SQD, several charge and energy transfer processes take place between these two materials. In such hybrid, it is known that when carriers are excited in SQD it can get transported to the neighboring MNP\cite{Ultrafasr_charge_Transfer_Samanta_2016,Counting_electron_Jayabalan_2019}. Any such carrier transport would result in quenching of the radiative emission and will show up as an increase in the amplitude of the fast decay component, $A_1$, along with the reduction in $A_2$ and $A_3$. Although this matches well to the observed kinetics (Table-\ref{Table1-nsPLdecay}) when excited at 2.25 eV, however at 3.06 eV excitation, we find a completely opposite situation where $A_1$ reduces with an increase in $A_2$ and $A_3$. Hence a simple charge transport model cannot consistently explain both the observed changes. In addition, several groups have also reported energy transfer between the metal and semiconductor components of the hybrid like foster resonance energy transfer (FRET) and plasmon induced resonance energy transfer (PIRET) \cite{ KANEMITSU2011510,Energy_transfer_Qd_Au_2004,LSPR_spectraloverlap_energy_transfer_dye_Au_2010,Size-Dependent_energy_transfer_CdSe_ZnSQD_Au_2011,Surface_energy_transfer_JPCC_2018}. In these process, energy get transferred non-radiatively from an emitter to an absorber by dipole-dipole coupling which occur when the emitter and absorber spectrally overlap\cite{KANEMITSU2011510,Energy_transfer_Qd_Au_2004,LSPR_spectraloverlap_energy_transfer_dye_Au_2010,Size-Dependent_energy_transfer_CdSe_ZnSQD_Au_2011,Surface_energy_transfer_JPCC_2018}. Similar to the charge transport case, these processes would also result in increasing $A_1$ with the reduction of $A_2$ and $A_3$. Once again these processes also could not explain the observed kinetics at 3.06 eV excitation. Further, as mentioned earlier, the presence of MNP would also create additional defect states at the surface of  SQD\cite{Nanoplasmonics_enhancement_Nanotechnology_2006,LSPR_Bisosensing_Ruemmele_ACS_2010,Plasmon_solar_energy_Scott_2013, EMAM2018287}. The excited charge carriers in SQD can get transported to these defect states and can relax further down radiatively. Such relaxation would result in reduction of $A_2$, the amplitude of radiative recombination, coupled with a PL emission having long lifetime. Such changes are also not observed in the present case. Thus all these reported process cannot consistently explain our observations. 

\begin{figure}[h] 
	\centerline{\includegraphics[width=0.9\columnwidth]{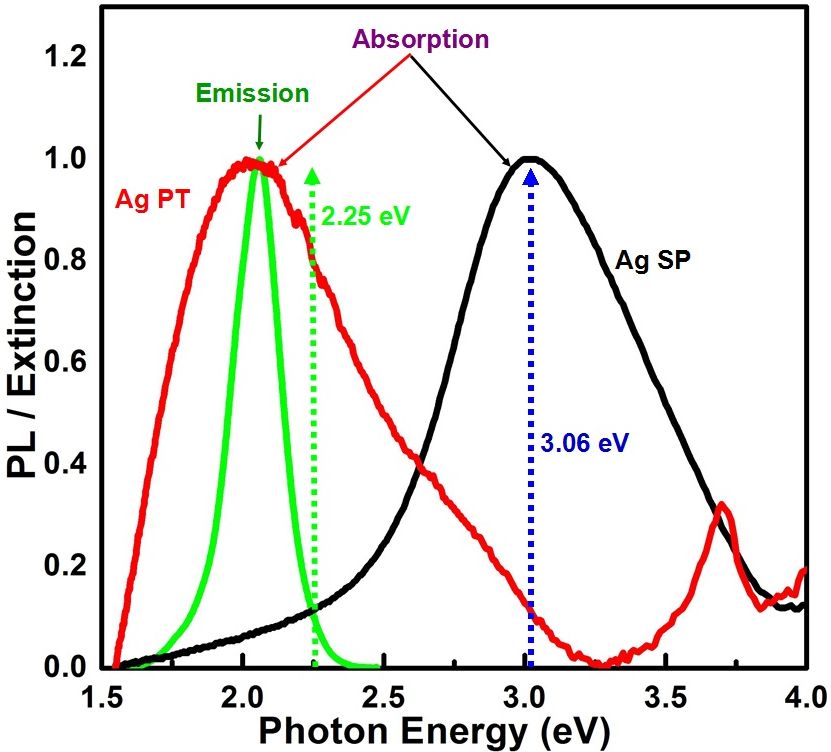}}
	\caption{Photoluminescence of CdTe QD colloid (green line) and absorption spectra of Ag platelet (red line) and Ag NS (black line) colloids. \label{Spectral-overlap}}
\end{figure}

In the present case, the aspect ratio of Ag PT is chosen, such that its in-plane dipole LSPR peak overlaps with that of CdTe QD emission wavelength (Fig.\ref{Spectral-overlap}). When H$_{AgPT}$ is excited at 3.06 eV, the LSPR does not get excited in Ag PT, however, the photon energy is sufficient to excite carriers in the CdTe QD. Ultrafast pump-probe measurements on such Ag nanoparticle-CdTe QD hybrid systems showed that electrons excited deep into the conduction band can get transported to Ag nanoparticles in subpicosecond and picosecond time scales creating a charge imbalance between them\cite{Ultrafasr_charge_Transfer_Samanta_2016,Counting_electron_Jayabalan_2019}. These carriers which are now in Ag nanoparticle returns to CdTe QD through defect states in few picoseconds time scale. In addition, the electrons in the CdTe QD can also relax to the newly formed defect states from which they can decay further down. In both of these cases, the presence of Ag PT will quickly reduce the number of carriers in the CdTe QD. Such a reduction in the number of charge carriers in CdTe QD strongly reduce the efficiency of the Auger recombination reducing the amplitude of the fast lifetime component, $A_{1}$. The reduction in the efficiency of Auger recombination can increase the PL emission from the band to band recombination\cite{LED_MRSbulletin_2013}. Spectral overlap between PL emission wavelength and LSPR peak (Fig.\ref{Spectral-overlap}) can also enhance the PL emission strength of CdTe QD in presence of MNP in a way like antenna effect \cite{PL_decay_CdSe_Au_nanoanteena_Haridas_2013,Nanoanteena_Silicon_Au_Kavokin_2019}. We attribute the increase in $A_2$ to the antenna effect enhancement and band to band recombination due to the reduction in Auger recombination. The carriers which are transported to the defect states will also relax radiatively increasing the emission from defect states\cite{Ultrafasr_charge_Transfer_Samanta_2016,Surface_related_emission_CdSe_Wang_2003}. The increase in $A_3$, the amplitude of the long time constant ($\tau_{3}$) can be attributed to the radiative relaxation through the defect states. Together with both of these processes, the observed change in the PL kinetics of H$_{AgPT}$ with respect to CdTe QD colloid could be explained. Figure\ref{Schematic}({\it b}) shows the schematic representation of different processes of PL kinetics in H$_{AgPT}$ when excited at 3.06 eV.

When the H$_{AgPT}$ hybrid sample is excited at 2.25 eV, charge carriers are still excited in CdTe QDs and simultaneously LSPR of Ag PT is also excited (Fig.\ref{Spectral-overlap}). Such simultaneous excitation can lead to mainly three different processes which are as follows. First the excitation of LSPR can result in enhancement of local field increasing the excited carrier density in CdTe QD\cite{Multiphoton_enhancement_CdSe_Au_film_Moyer_2013,SinglePhoton_emission_JPCC_2011}. Secondly the plasmon can relax by exciting carriers in CdTe QD leading to a PIRET\cite{CdSe_Au_nanorod_Focsan_2014,Multiexciton_emission_Zhao_2015}. Third, the hot electrons that are formed after plasmon relaxation can also get transferred to CdTe QD in subpicosecond time scale\cite{Counting_electron_Jayabalan_2019}. All these three processes will result in strong increase in the excited carrier density in CdTe QD. Such increase in the density of carriers will actively facilitate an increase in Auger recombination, increasing the $A_1$ and further reducing $\tau_{1}$. In addition to this, increase in the Auger recombination will also reduce the strength of radiative recombination  ($A_2$) as well as reduces the rate of transport of carriers to the defect states, reducing $A_3$. This explains all the changes observed in the experiment (Table-\ref{Table1-nsPLdecay}). Figure\ref{Schematic}({\it c}) shows the schematic representation of different processes of PL kinetics in H$_{AgPT}$ when excited at 2.25 eV. Thus, the excitation photon energy with respect to the overlapped emission spectrum and LSPR absorption could alter the PL kinetics drastically in Ag-CdTe HNS. In a hybrid system where these two would not overlap such changes in the  PL kinetics cannot be observed and would lead to overall quenching of PL.

To verify the proposed model, we performed similar measurement on Ag nanosphere (SP)-CdTe QD hybrid system, (H$_{AgSP}$), where PL emission \& LSPR do not overlap. The details of the preparation and structural \& optical characterization of Ag SP and H$_{AgSP}$ are given in the SI(Fig.S2 and Fig.S3). The CdTe QD used in the preparation of H$_{AgSP}$ colloid is same as that used in H$_{AgPT}$ colloid. In Figure\ref{Spectral-overlap}, we also show the absorption spectrum of the SP (Ag SP) colloid used for the preparation of H$_{AgSP}$. In this hybrid, the emission spectrum of CdTe QD which has peak at 2 eV, does not overlap with the LSPR of Ag SP which is at 3.02 eV. In Figure\ref{405nm_lifetime} and \ref{550nm_lifetime} we also show the time dependence of the H$_{AgSP}$ when excited at 3.06 eV and 2.25 eV respectively. The best-fit parameters obtained by fitting the time-dependent PL is also summarized in Table-\ref{Table1-nsPLdecay}. Unlike the H$_{AgPT}$ case, this time, there is not much difference in the PL dynamics when the excitation energy is changed from 3.06 eV to 2.25 eV. Excitation at 3.06 eV excites carriers in the CdTe QD and simulatensouly LSPR also gets excited. As discussed earlier, such excitation would create an increase in the carrier density in CdTe QD which results in enhancing Auger recombination. The observed increase in $A_1$ with simultaneous reduction in $A_2$ and $A_2$, the amplitudes corresponding to radiative and defect state emissions, matches well with the proposed model. When excited at 2.25 eV, LSPR is not excited still hot carriers can get transported to Ag SP in few hundreds of femtosecond and few picosecond time scales\cite{Ultrafasr_charge_Transfer_Samanta_2016}. However, unlike the H$_{AgPT}$ where presence of MNP can enhance the radiative emission, in case of H$_{AgSP}$ the hot carrier transport results only in fast quenching of PL, reducing $A_2$ and $A_3$. Therefore the total integrated area under the PL spectrum quenches to 70\% for both 3.06 eV and 2.25 eV excitation compared to that of corresponding CdTe QD colloid(Fig.S3). Thus the PL kinetics of Ag-CdTe hybrid can be controlled by selective excitation only when the PL emission of SQD overlaps with LSPR absorption of MNP.

\section {Conclusion}
Methods like doping, forming core-shell structures and alloying are being studied to suppress auger recombination and to control the PL kinetics in QDs. Our studies on the excitation photon energy dependence of PL kinetics in suitably designed Ag-CdTe hybrids shows that it is possible to selectively excite and drastically alter the PL kinetics. Excitation of a hybrid system well above an overlapped emission (of SQD) and LSPR (of MNP) could effectively suppress the Auger recombination. On the other hand, exciting right at the overlapped spectral regime enhances the Auger recombination. By suitably tuning the LSPR of MNP it is possible to assemble a metal-semiconductor hybrid which can show suitable PL kinetics at a given excitation energy. Controlling the Auger recombination in plasmon controlled PL kinetics  in SQD  by placing a MNP near to it opens up a new method of designing a material with efficient radiative recombination that is well suited for light-emitting devices. 

\section{Acknowledgement}
The authors are thankful to Tarun Kumar Sharma and Rama Chari for fruitful discussions and suggestions. The author, Sabina Gurung, is thankful to RRCAT, Indore, under HBNI program, Mumbai, for the financial support.

\newpage 
\bibliographystyle{rsc} 
\bibliography{Article-001}

\providecommand*{\mcitethebibliography}{\thebibliography}
\csname @ifundefined\endcsname{endmcitethebibliography}
{\let\endmcitethebibliography\endthebibliography}{}
\begin{mcitethebibliography}{56}
\providecommand*{\natexlab}[1]{#1}
\providecommand*{\mciteSetBstSublistMode}[1]{}
\providecommand*{\mciteSetBstMaxWidthForm}[2]{}
\providecommand*{\mciteBstWouldAddEndPuncttrue}
  {\def\EndOfBibitem{\unskip.}}
\providecommand*{\mciteBstWouldAddEndPunctfalse}
  {\let\EndOfBibitem\relax}
\providecommand*{\mciteSetBstMidEndSepPunct}[3]{}
\providecommand*{\mciteSetBstSublistLabelBeginEnd}[3]{}
\providecommand*{\EndOfBibitem}{}
\mciteSetBstSublistMode{f}
\mciteSetBstMaxWidthForm{subitem}
{(\emph{\alph{mcitesubitemcount}})}
\mciteSetBstSublistLabelBeginEnd{\mcitemaxwidthsubitemform\space}
{\relax}{\relax}

\bibitem[Wuister \emph{et~al.}(2003)Wuister, Swart, van Driel, Hickey, and
  de~Mello~Donegá]{Luminescent_CdTe_QD_2003}
S.~F. Wuister, I.~Swart, F.~van Driel, S.~G. Hickey and C.~de~Mello~Donegá,
  \emph{Nano Letters}, 2003, \textbf{3}, 503--507\relax
\mciteBstWouldAddEndPuncttrue
\mciteSetBstMidEndSepPunct{\mcitedefaultmidpunct}
{\mcitedefaultendpunct}{\mcitedefaultseppunct}\relax
\EndOfBibitem
\bibitem[Rakovich \emph{et~al.}(2003)Rakovich, Walsh, Bradley, Donegan,
  Talapin, Rogach, and Eychmueller]{Size_selective_PL_CdTeQD_2003}
Y.~Rakovich, L.~Walsh, L.~Bradley, J.~F. Donegan, D.~Talapin, A.~Rogach and
  A.~Eychmueller, Opto-Ireland 2002: Optics and Photonics Technologies and
  Applications, 2003, pp. 432 -- 437\relax
\mciteBstWouldAddEndPuncttrue
\mciteSetBstMidEndSepPunct{\mcitedefaultmidpunct}
{\mcitedefaultendpunct}{\mcitedefaultseppunct}\relax
\EndOfBibitem
\bibitem[Guo \emph{et~al.}(2005)Guo, Yang, and
  Wang]{Guo-ReactionConditions-TGA-CdTeNP-JPCB-2005}
J.~Guo, W.~Yang and C.~Wang, \emph{J. Phys. Chem. B}, 2005, \textbf{109},
  17467--17473\relax
\mciteBstWouldAddEndPuncttrue
\mciteSetBstMidEndSepPunct{\mcitedefaultmidpunct}
{\mcitedefaultendpunct}{\mcitedefaultseppunct}\relax
\EndOfBibitem
\bibitem[Tang and Ouyang(2007)]{Tailoring_Properties_crystanality_Nature_2007}
Y.~Tang and M.~Ouyang, \emph{Nat. Mater.}, 2007, \textbf{6}, 754\relax
\mciteBstWouldAddEndPuncttrue
\mciteSetBstMidEndSepPunct{\mcitedefaultmidpunct}
{\mcitedefaultendpunct}{\mcitedefaultseppunct}\relax
\EndOfBibitem
\bibitem[Haldar \emph{et~al.}(2012)Haldar, Sen, Mandal, and
  Patra]{photophysical_properties_Au_CdTe_varying_size_shape_2012}
K.~K. Haldar, T.~Sen, S.~Mandal and A.~Patra, \emph{ChemPhysChem}, 2012,
  \textbf{13}, 3989--3996\relax
\mciteBstWouldAddEndPuncttrue
\mciteSetBstMidEndSepPunct{\mcitedefaultmidpunct}
{\mcitedefaultendpunct}{\mcitedefaultseppunct}\relax
\EndOfBibitem
\bibitem[Viste \emph{et~al.}(2010)Viste, Plain, Jaffiol, Vial, Adam, and
  Royer]{Enhancement_quenching_metal_sc_hybrid_Viste_2010}
P.~Viste, J.~Plain, R.~Jaffiol, A.~Vial, P.~M. Adam and P.~Royer, \emph{ACS
  Nano}, 2010, \textbf{4}, 759--764\relax
\mciteBstWouldAddEndPuncttrue
\mciteSetBstMidEndSepPunct{\mcitedefaultmidpunct}
{\mcitedefaultendpunct}{\mcitedefaultseppunct}\relax
\EndOfBibitem
\bibitem[Jiang \emph{et~al.}(2014)Jiang, Li, Fang, and
  Wang]{metal_sc_plasmon_enhanced_Jiang_Adv_matter_2014}
R.~Jiang, B.~Li, C.~Fang and J.~Wang, \emph{Adv. Mater.}, 2014, \textbf{26},
  5274--5309\relax
\mciteBstWouldAddEndPuncttrue
\mciteSetBstMidEndSepPunct{\mcitedefaultmidpunct}
{\mcitedefaultendpunct}{\mcitedefaultseppunct}\relax
\EndOfBibitem
\bibitem[Strelow \emph{et~al.}(2016)Strelow, Theuerholz, Schmidtke, Richter,
  Merkl, Kloust, Ye, Weller, Heinz,
  Knorr,\emph{et~al.}]{metal_Sc_hybrid_selforganization_exciton_plsmon_Strelow_2016}
C.~Strelow, T.~S. Theuerholz, C.~Schmidtke, M.~Richter, J.-P. Merkl, H.~Kloust,
  Z.~Ye, H.~Weller, T.~F. Heinz, A.~Knorr \emph{et~al.}, \emph{Nano Lett.},
  2016, \textbf{16}, 4811--4818\relax
\mciteBstWouldAddEndPuncttrue
\mciteSetBstMidEndSepPunct{\mcitedefaultmidpunct}
{\mcitedefaultendpunct}{\mcitedefaultseppunct}\relax
\EndOfBibitem
\bibitem[Mie(1908)]{Gustav_Mie_LSPR_1908}
G.~Mie, \emph{Annalen der Physik}, 1908, \textbf{330}, 377--445\relax
\mciteBstWouldAddEndPuncttrue
\mciteSetBstMidEndSepPunct{\mcitedefaultmidpunct}
{\mcitedefaultendpunct}{\mcitedefaultseppunct}\relax
\EndOfBibitem
\bibitem[Petryayeva and Krull(2011)]{LSPR_review_Analytica_chimica_2011}
E.~Petryayeva and U.~J. Krull, \emph{Anal. Chim. Acta}, 2011, \textbf{706},
  8--24\relax
\mciteBstWouldAddEndPuncttrue
\mciteSetBstMidEndSepPunct{\mcitedefaultmidpunct}
{\mcitedefaultendpunct}{\mcitedefaultseppunct}\relax
\EndOfBibitem
\bibitem[Wuister \emph{et~al.}(2004)Wuister, de~Mello~Donegá, and
  Meijerink]{Local_field_spontaneous_rate_CdTe_CdSe_2004}
S.~F. Wuister, C.~de~Mello~Donegá and A.~Meijerink, \emph{The Journal of
  Chemical Physics}, 2004, \textbf{121}, 4310--4315\relax
\mciteBstWouldAddEndPuncttrue
\mciteSetBstMidEndSepPunct{\mcitedefaultmidpunct}
{\mcitedefaultendpunct}{\mcitedefaultseppunct}\relax
\EndOfBibitem
\bibitem[Ruemmele \emph{et~al.}(2013)Ruemmele, Hall, Ruvuna, and
  Van~Duyne]{LSPR_Bisosensing_Ruemmele_ACS_2010}
J.~A. Ruemmele, W.~P. Hall, L.~K. Ruvuna and R.~P. Van~Duyne, \emph{Anal.
  Chem.}, 2013, \textbf{85}, 4560--4566\relax
\mciteBstWouldAddEndPuncttrue
\mciteSetBstMidEndSepPunct{\mcitedefaultmidpunct}
{\mcitedefaultendpunct}{\mcitedefaultseppunct}\relax
\EndOfBibitem
\bibitem[Bharadwaj \emph{et~al.}(2006)Bharadwaj, Anger, and
  Novotny]{Nanoplasmonics_enhancement_Nanotechnology_2006}
P.~Bharadwaj, P.~Anger and L.~Novotny, \emph{Nanotechnology}, 2006,
  \textbf{18}, 044017\relax
\mciteBstWouldAddEndPuncttrue
\mciteSetBstMidEndSepPunct{\mcitedefaultmidpunct}
{\mcitedefaultendpunct}{\mcitedefaultseppunct}\relax
\EndOfBibitem
\bibitem[Cushing and Wu(2013)]{Plasmon_solar_energy_Scott_2013}
S.~K. Cushing and N.~Wu, \emph{Electrochem. Soc. Interface}, 2013, \textbf{22},
  63--67\relax
\mciteBstWouldAddEndPuncttrue
\mciteSetBstMidEndSepPunct{\mcitedefaultmidpunct}
{\mcitedefaultendpunct}{\mcitedefaultseppunct}\relax
\EndOfBibitem
\bibitem[Emam \emph{et~al.}(2018)Emam, Mostafa, Mohamed, Gadallah, and
  El-Kemary]{EMAM2018287}
A.~N. Emam, A.~A. Mostafa, M.~B. Mohamed, A.-S. Gadallah and M.~El-Kemary,
  \emph{Journal of Luminescence}, 2018, \textbf{200}, 287 -- 297\relax
\mciteBstWouldAddEndPuncttrue
\mciteSetBstMidEndSepPunct{\mcitedefaultmidpunct}
{\mcitedefaultendpunct}{\mcitedefaultseppunct}\relax
\EndOfBibitem
\bibitem[Wang \emph{et~al.}(2006)Wang, Li, Jia, Song, Han, Zhang, Yang, Xu, and
  Zhao]{Ag_cdte_selforganized_electrostatic_interaction_Wang_Spect_Acta_2005}
Y.~Wang, M.~Li, H.~Jia, W.~Song, X.~Han, J.~Zhang, B.~Yang, W.~Xu and B.~Zhao,
  \emph{Spectrochim. Acta, Part A}, 2006, \textbf{64}, 101--105\relax
\mciteBstWouldAddEndPuncttrue
\mciteSetBstMidEndSepPunct{\mcitedefaultmidpunct}
{\mcitedefaultendpunct}{\mcitedefaultseppunct}\relax
\EndOfBibitem
\bibitem[Mondal and Samanta(2016)]{Ultrafasr_charge_Transfer_Samanta_2016}
N.~Mondal and A.~Samanta, \emph{The Journal of Physical Chemistry C}, 2016,
  \textbf{120}, 650--658\relax
\mciteBstWouldAddEndPuncttrue
\mciteSetBstMidEndSepPunct{\mcitedefaultmidpunct}
{\mcitedefaultendpunct}{\mcitedefaultseppunct}\relax
\EndOfBibitem
\bibitem[Gurung \emph{et~al.}(2018)Gurung, Singh, Chari, and
  Jayabalan]{Sabina-JAP-2018}
S.~Gurung, A.~Singh, R.~Chari and J.~Jayabalan, \emph{J. Appl. Phys.}, 2018,
  \textbf{124}, 204305\relax
\mciteBstWouldAddEndPuncttrue
\mciteSetBstMidEndSepPunct{\mcitedefaultmidpunct}
{\mcitedefaultendpunct}{\mcitedefaultseppunct}\relax
\EndOfBibitem
\bibitem[Cushing \emph{et~al.}(2015)Cushing, Li, Bright, Yost, Zheng, Bristow,
  and Wu]{Controlling_PIRET_Metal_TiO2_JPCC_2015}
S.~K. Cushing, J.~Li, J.~Bright, B.~T. Yost, P.~Zheng, A.~D. Bristow and N.~Wu,
  \emph{The Journal of Physical Chemistry C}, 2015, \textbf{119},
  16239--16244\relax
\mciteBstWouldAddEndPuncttrue
\mciteSetBstMidEndSepPunct{\mcitedefaultmidpunct}
{\mcitedefaultendpunct}{\mcitedefaultseppunct}\relax
\EndOfBibitem
\bibitem[Wargnier \emph{et~al.}(2004)Wargnier, Baranov, Maslov, Stsiapura,
  Artemyev, Pluot, Sukhanova, and Nabiev]{Energy_transfer_Qd_Au_2004}
R.~Wargnier, A.~V. Baranov, V.~G. Maslov, V.~Stsiapura, M.~Artemyev, M.~Pluot,
  A.~Sukhanova and I.~Nabiev, \emph{Nano Lett.}, 2004, \textbf{4},
  4451--457\relax
\mciteBstWouldAddEndPuncttrue
\mciteSetBstMidEndSepPunct{\mcitedefaultmidpunct}
{\mcitedefaultendpunct}{\mcitedefaultseppunct}\relax
\EndOfBibitem
\bibitem[Li \emph{et~al.}(2011)Li, Cushing, Wang, Shi, Horna, Hong, and
  Wu]{Size-Dependent_energy_transfer_CdSe_ZnSQD_Au_2011}
M.~Li, S.~K. Cushing, Q.~Wang, X.~Shi, L.~A. Horna, Z.~Hong and N.~Wu,
  \emph{J.Phys.Chem.Lett.}, 2011, \textbf{2}, 2125--2129\relax
\mciteBstWouldAddEndPuncttrue
\mciteSetBstMidEndSepPunct{\mcitedefaultmidpunct}
{\mcitedefaultendpunct}{\mcitedefaultseppunct}\relax
\EndOfBibitem
\bibitem[Kanemitsu and Matsuda(2011)]{KANEMITSU2011510}
Y.~Kanemitsu and K.~Matsuda, \emph{Journal of Luminescence}, 2011,
  \textbf{131}, 510 -- 514\relax
\mciteBstWouldAddEndPuncttrue
\mciteSetBstMidEndSepPunct{\mcitedefaultmidpunct}
{\mcitedefaultendpunct}{\mcitedefaultseppunct}\relax
\EndOfBibitem
\bibitem[Vaishnav and Mukherjee(2018)]{Surface_energy_transfer_JPCC_2018}
J.~K. Vaishnav and T.~K. Mukherjee, \emph{J. Phys. Chem. C}, 2018,
  \textbf{122}, 28324--28336\relax
\mciteBstWouldAddEndPuncttrue
\mciteSetBstMidEndSepPunct{\mcitedefaultmidpunct}
{\mcitedefaultendpunct}{\mcitedefaultseppunct}\relax
\EndOfBibitem
\bibitem[Singh and
  Strouse(2010)]{LSPR_spectraloverlap_energy_transfer_dye_Au_2010}
M.~P. Singh and G.~F. Strouse, \emph{Journal of the American Chemical Society},
  2010, \textbf{132}, 9383--9391\relax
\mciteBstWouldAddEndPuncttrue
\mciteSetBstMidEndSepPunct{\mcitedefaultmidpunct}
{\mcitedefaultendpunct}{\mcitedefaultseppunct}\relax
\EndOfBibitem
\bibitem[Gao \emph{et~al.}(2012)Gao, Lin, Wei, Zeng, Liao, Chen, Goldfeld,
  Wang, Luo, Dong, and Hou ]{Charge_transfer_Au_CdSe_Gao_2012}
B.~Gao, Y.~Lin, S.~Wei, J.~Zeng, Y.~Liao, L.~Chen, D.~Goldfeld, X.~Wang,
  Y.~Luo, Z.~Dong and J.~Hou , \emph{Nano Research}, 2012, \textbf{5},
  88--98\relax
\mciteBstWouldAddEndPuncttrue
\mciteSetBstMidEndSepPunct{\mcitedefaultmidpunct}
{\mcitedefaultendpunct}{\mcitedefaultseppunct}\relax
\EndOfBibitem
\bibitem[Dey and Zhao(2016)]{Plasmon_effect_multiexciton_QD_Zhao_2016}
S.~Dey and J.~Zhao, \emph{The Journal of Physical Chemistry Letters}, 2016,
  \textbf{7}, 2921--2929\relax
\mciteBstWouldAddEndPuncttrue
\mciteSetBstMidEndSepPunct{\mcitedefaultmidpunct}
{\mcitedefaultendpunct}{\mcitedefaultseppunct}\relax
\EndOfBibitem
\bibitem[LeBlanc \emph{et~al.}(2013)LeBlanc, McClanahan, Jones, and
  Moyer]{Multiphoton_enhancement_CdSe_Au_film_Moyer_2013}
S.~J. LeBlanc, M.~R. McClanahan, M.~Jones and P.~J. Moyer, \emph{Nano Letters},
  2013, \textbf{13}, 1662--1669\relax
\mciteBstWouldAddEndPuncttrue
\mciteSetBstMidEndSepPunct{\mcitedefaultmidpunct}
{\mcitedefaultendpunct}{\mcitedefaultseppunct}\relax
\EndOfBibitem
\bibitem[Bae \emph{et~al.}(2013)Bae, Brovelli, and
  Klimov]{LED_MRSbulletin_2013}
W.~K. Bae, S.~Brovelli and V.~I. Klimov, \emph{MRS Bulletin}, 2013,
  \textbf{38}, 721–730\relax
\mciteBstWouldAddEndPuncttrue
\mciteSetBstMidEndSepPunct{\mcitedefaultmidpunct}
{\mcitedefaultendpunct}{\mcitedefaultseppunct}\relax
\EndOfBibitem
\bibitem[Jin \emph{et~al.}(2001)Jin, Cao, Mirkin, Kelly, Schatz, and
  Zheng]{Silver_sphere_Science_2001}
R.~Jin, Y.~Cao, C.~A. Mirkin, K.~Kelly, G.~C. Schatz and J.~Zheng,
  \emph{Science}, 2001, \textbf{294}, 1901--1903\relax
\mciteBstWouldAddEndPuncttrue
\mciteSetBstMidEndSepPunct{\mcitedefaultmidpunct}
{\mcitedefaultendpunct}{\mcitedefaultseppunct}\relax
\EndOfBibitem
\bibitem[Mandal and Tamai(2008)]{Abhijit-CdTe-Pre-PL-JPCC-2008}
A.~Mandal and N.~Tamai, \emph{J. Phys. Chem. C}, 2008, \textbf{112},
  8244--8250\relax
\mciteBstWouldAddEndPuncttrue
\mciteSetBstMidEndSepPunct{\mcitedefaultmidpunct}
{\mcitedefaultendpunct}{\mcitedefaultseppunct}\relax
\EndOfBibitem
\bibitem[Singh \emph{et~al.}(2019)Singh, Gurung, Chari, and
  Jayabalan]{Counting_electron_Jayabalan_2019}
A.~Singh, S.~Gurung, R.~Chari and J.~Jayabalan, \emph{The Journal of Physical
  Chemistry C}, 2019, \textbf{123}, 28584--28592\relax
\mciteBstWouldAddEndPuncttrue
\mciteSetBstMidEndSepPunct{\mcitedefaultmidpunct}
{\mcitedefaultendpunct}{\mcitedefaultseppunct}\relax
\EndOfBibitem
\bibitem[Yu \emph{et~al.}(2003)Yu, Qu, Guo, and
  Peng]{Pengs_cdte_quantum_dots_size_2003}
W.~W. Yu, L.~Qu, W.~Guo and X.~Peng, \emph{Chem. Mater.}, 2003, \textbf{15},
  2854--2860\relax
\mciteBstWouldAddEndPuncttrue
\mciteSetBstMidEndSepPunct{\mcitedefaultmidpunct}
{\mcitedefaultendpunct}{\mcitedefaultseppunct}\relax
\EndOfBibitem
\bibitem[Chen \emph{et~al.}(2005)Chen, Joly, and McCready]{UCPL_CdSe_2005}
W.~Chen, A.~G. Joly and D.~E. McCready, \emph{The Journal of Chemical Physics},
  2005, \textbf{122}, 224708\relax
\mciteBstWouldAddEndPuncttrue
\mciteSetBstMidEndSepPunct{\mcitedefaultmidpunct}
{\mcitedefaultendpunct}{\mcitedefaultseppunct}\relax
\EndOfBibitem
\bibitem[Jayabalan \emph{et~al.}(2010)Jayabalan, Singh, and
  Chari]{JJ-APL-Edgesmoothening}
J.~Jayabalan, A.~Singh and R.~Chari, \emph{Applied Physics Letters}, 2010,
  \textbf{97}, 041904\relax
\mciteBstWouldAddEndPuncttrue
\mciteSetBstMidEndSepPunct{\mcitedefaultmidpunct}
{\mcitedefaultendpunct}{\mcitedefaultseppunct}\relax
\EndOfBibitem
\bibitem[Chang \emph{et~al.}(2017)Chang, Ogomi, Ding, Zhang, Toyoda, Hayase,
  Katayama, and Shen]{Ligand_exciton_photovoltaic_PbS_Jin_PCCP_2017}
J.~Chang, Y.~Ogomi, C.~Ding, Y.~H. Zhang, T.~Toyoda, S.~Hayase, K.~Katayama and
  Q.~Shen, \emph{Phys. Chem. Chem. Phys.}, 2017, \textbf{19}, 6358--6367\relax
\mciteBstWouldAddEndPuncttrue
\mciteSetBstMidEndSepPunct{\mcitedefaultmidpunct}
{\mcitedefaultendpunct}{\mcitedefaultseppunct}\relax
\EndOfBibitem
\bibitem[Gallagher \emph{et~al.}(2013)Gallagher, Comby, Wojdyla, Gunnlaugsson,
  Kelly, Gun'ko, Clark, Greetham, Towrie, and
  Quinn]{TGA_CdTe_Europium_surface_coordinated_emission_Gallagher_inorganic_chemistry_2013}
S.~A. Gallagher, S.~Comby, M.~Wojdyla, T.~Gunnlaugsson, J.~M. Kelly, Y.~K.
  Gun'ko, I.~P. Clark, G.~M. Greetham, M.~Towrie and S.~J. Quinn, \emph{Inorg.
  Chem.}, 2013, \textbf{52}, 4133--4135\relax
\mciteBstWouldAddEndPuncttrue
\mciteSetBstMidEndSepPunct{\mcitedefaultmidpunct}
{\mcitedefaultendpunct}{\mcitedefaultseppunct}\relax
\EndOfBibitem
\bibitem[Govorov \emph{et~al.}(2006)Govorov, Bryant, Zhang, Skeini, Lee, Kotov,
  Slocik, and Naik]{Govorov_NanoLett_2006}
A.~O. Govorov, G.~W. Bryant, W.~Zhang, T.~Skeini, J.~Lee, N.~A. Kotov, J.~M.
  Slocik and R.~R. Naik, \emph{Nano Letters}, 2006, \textbf{6}, 984--994\relax
\mciteBstWouldAddEndPuncttrue
\mciteSetBstMidEndSepPunct{\mcitedefaultmidpunct}
{\mcitedefaultendpunct}{\mcitedefaultseppunct}\relax
\EndOfBibitem
\bibitem[Kim \emph{et~al.}(2013)Kim, Yokota, Taniguchi, and
  Nakayama]{Precise_control_JAP_2013}
D.~Kim, H.~Yokota, T.~Taniguchi and M.~Nakayama, \emph{Journal of Applied
  Physics}, 2013, \textbf{114}, 154307\relax
\mciteBstWouldAddEndPuncttrue
\mciteSetBstMidEndSepPunct{\mcitedefaultmidpunct}
{\mcitedefaultendpunct}{\mcitedefaultseppunct}\relax
\EndOfBibitem
\bibitem[Viste \emph{et~al.}(2010)Viste, Plain, Jaffiol, Vial, Adam, and
  Royer]{Viste_ACSNano_2010}
P.~Viste, J.~Plain, R.~Jaffiol, A.~Vial, P.~M. Adam and P.~Royer, \emph{ACS
  Nano}, 2010, \textbf{4}, 759--764\relax
\mciteBstWouldAddEndPuncttrue
\mciteSetBstMidEndSepPunct{\mcitedefaultmidpunct}
{\mcitedefaultendpunct}{\mcitedefaultseppunct}\relax
\EndOfBibitem
\bibitem[Xia \emph{et~al.}(2008)Xia, Cao, and Zhu]{XIA2008166}
Y.-S. Xia, C.~Cao and C.-Q. Zhu, \emph{Journal of Luminescence}, 2008,
  \textbf{128}, 166 -- 172\relax
\mciteBstWouldAddEndPuncttrue
\mciteSetBstMidEndSepPunct{\mcitedefaultmidpunct}
{\mcitedefaultendpunct}{\mcitedefaultseppunct}\relax
\EndOfBibitem
\bibitem[An \emph{et~al.}(2010)An, Yang, Su, Yi, Liu, Chao, and
  Zeng]{Enhanced_PL_CdTe_Ag}
L.~M. An, Y.~Q. Yang, W.~H. Su, J.~Yi, C.~X. Liu, K.~F. Chao and Q.~H. Zeng,
  \emph{Journal of Nanoscience and Nanotechnology}, 2010, \textbf{10},
  2099--2103\relax
\mciteBstWouldAddEndPuncttrue
\mciteSetBstMidEndSepPunct{\mcitedefaultmidpunct}
{\mcitedefaultendpunct}{\mcitedefaultseppunct}\relax
\EndOfBibitem
\bibitem[Nahar \emph{et~al.}(2015)Nahar, Esteves, Hafiz, Ozgur, and
  Arachchige]{Nahar_ACSNano_2015}
L.~Nahar, R.~J.~A. Esteves, S.~Hafiz, U.~Ozgur and I.~U. Arachchige, \emph{ACS
  Nano}, 2015, \textbf{9}, 9810--9821\relax
\mciteBstWouldAddEndPuncttrue
\mciteSetBstMidEndSepPunct{\mcitedefaultmidpunct}
{\mcitedefaultendpunct}{\mcitedefaultseppunct}\relax
\EndOfBibitem
\bibitem[Wang \emph{et~al.}(2003)Wang, Qu, Zhang, Peng, and
  Xiao]{Surface_related_emission_CdSe_Wang_2003}
X.~Wang, L.~Qu, J.~Zhang, X.~Peng and M.~Xiao, \emph{Nano Letters}, 2003,
  \textbf{3}, 1103--1106\relax
\mciteBstWouldAddEndPuncttrue
\mciteSetBstMidEndSepPunct{\mcitedefaultmidpunct}
{\mcitedefaultendpunct}{\mcitedefaultseppunct}\relax
\EndOfBibitem
\bibitem[Rogach \emph{et~al.}(2007)Rogach, Franzl, Klar, Feldmann, Gaponik,
  Lesnyak, Shavel, Eychmüller, Rakovich, and Donegan]{TGA_CdTe_2007}
A.~L. Rogach, T.~Franzl, T.~A. Klar, J.~Feldmann, N.~Gaponik, V.~Lesnyak,
  A.~Shavel, A.~Eychmüller, Y.~P. Rakovich and J.~F. Donegan, \emph{The
  Journal of Physical Chemistry C}, 2007, \textbf{111}, 14628--14637\relax
\mciteBstWouldAddEndPuncttrue
\mciteSetBstMidEndSepPunct{\mcitedefaultmidpunct}
{\mcitedefaultendpunct}{\mcitedefaultseppunct}\relax
\EndOfBibitem
\bibitem[Mandal and Tamai(2008)]{InfluenceofAcid_PL_TGA_CdTe_Mandal_2008}
A.~Mandal and N.~Tamai, \emph{The Journal of Physical Chemistry C}, 2008,
  \textbf{112}, 8244--8250\relax
\mciteBstWouldAddEndPuncttrue
\mciteSetBstMidEndSepPunct{\mcitedefaultmidpunct}
{\mcitedefaultendpunct}{\mcitedefaultseppunct}\relax
\EndOfBibitem
\bibitem[Espinobarro-Velazquez \emph{et~al.}(2015)Espinobarro-Velazquez,
  Leontiadou, Page, Califano, O'Brien, and
  Binks]{Effect_Chloride_Passivation_CdTe_Binks_2015}
D.~Espinobarro-Velazquez, M.~A. Leontiadou, R.~C. Page, M.~Califano, P.~O'Brien
  and D.~J. Binks, \emph{ChemPhysChem}, 2015, \textbf{16}, 1239--1244\relax
\mciteBstWouldAddEndPuncttrue
\mciteSetBstMidEndSepPunct{\mcitedefaultmidpunct}
{\mcitedefaultendpunct}{\mcitedefaultseppunct}\relax
\EndOfBibitem
\bibitem[Dey \emph{et~al.}(2018)Dey, Zhou, Sun, Jenkins, Kriz, Suib, Chen, Zou,
  and Zhao]{Antibunched_QD_Zhao_2018}
S.~Dey, Y.~Zhou, Y.~Sun, J.~A. Jenkins, D.~Kriz, S.~L. Suib, O.~Chen, S.~Zou
  and J.~Zhao, \emph{Nanoscale}, 2018, \textbf{10}, 1038--1046\relax
\mciteBstWouldAddEndPuncttrue
\mciteSetBstMidEndSepPunct{\mcitedefaultmidpunct}
{\mcitedefaultendpunct}{\mcitedefaultseppunct}\relax
\EndOfBibitem
\bibitem[Sagarzazu \emph{et~al.}(2011)Sagarzazu, Kobayashi, Murase, Yang, and
  Tamai]{Auger_CdTe_PCCP_2013}
G.~Sagarzazu, Y.~Kobayashi, N.~Murase, P.~Yang and N.~Tamai, \emph{Phys. Chem.
  Chem. Phys.}, 2011, \textbf{13}, 3227--3230\relax
\mciteBstWouldAddEndPuncttrue
\mciteSetBstMidEndSepPunct{\mcitedefaultmidpunct}
{\mcitedefaultendpunct}{\mcitedefaultseppunct}\relax
\EndOfBibitem
\bibitem[Kobayashi \emph{et~al.}(2011)Kobayashi, Nishimura, Yamaguchi, and
  Tamai]{Auger_CdS_JPCL_2011}
Y.~Kobayashi, T.~Nishimura, H.~Yamaguchi and N.~Tamai, \emph{The Journal of
  Physical Chemistry Letters}, 2011, \textbf{2}, 1051--1055\relax
\mciteBstWouldAddEndPuncttrue
\mciteSetBstMidEndSepPunct{\mcitedefaultmidpunct}
{\mcitedefaultendpunct}{\mcitedefaultseppunct}\relax
\EndOfBibitem
\bibitem[Wax \emph{et~al.}(2018)Wax, Dey, Chen, Luo, Zou, and
  Zhao]{Excitation_depend_PL_decay_Au_QD_Zhao_2018}
T.~J. Wax, S.~Dey, S.~Chen, Y.~Luo, S.~Zou and J.~Zhao, \emph{ACS Omega}, 2018,
  \textbf{3}, 14151--14156\relax
\mciteBstWouldAddEndPuncttrue
\mciteSetBstMidEndSepPunct{\mcitedefaultmidpunct}
{\mcitedefaultendpunct}{\mcitedefaultseppunct}\relax
\EndOfBibitem
\bibitem[Wang \emph{et~al.}(2003)Wang, Yu, Zhang, Aldana, Peng, and
  Xiao]{PL_Upconversion_CdTe_Xiao_2003}
X.~Wang, W.~W. Yu, J.~Zhang, J.~Aldana, X.~Peng and M.~Xiao, \emph{Phys. Rev.
  B}, 2003, \textbf{68}, 125318\relax
\mciteBstWouldAddEndPuncttrue
\mciteSetBstMidEndSepPunct{\mcitedefaultmidpunct}
{\mcitedefaultendpunct}{\mcitedefaultseppunct}\relax
\EndOfBibitem
\bibitem[Haridas \emph{et~al.}(2013)Haridas, Basu, Tiwari, and
  Venkatapathi]{PL_decay_CdSe_Au_nanoanteena_Haridas_2013}
M.~Haridas, J.~K. Basu, A.~K. Tiwari and M.~Venkatapathi, \emph{Journal of
  Applied Physics}, 2013, \textbf{114}, 064305\relax
\mciteBstWouldAddEndPuncttrue
\mciteSetBstMidEndSepPunct{\mcitedefaultmidpunct}
{\mcitedefaultendpunct}{\mcitedefaultseppunct}\relax
\EndOfBibitem
\bibitem[Kucherik \emph{et~al.}(2019)Kucherik, Kutrovskaya, Osipov, Gerke,
  Chestnov, Arakelian, Shalin, Evlyukhin, and
  Kavokin ]{Nanoanteena_Silicon_Au_Kavokin_2019}
A.~Kucherik, S.~Kutrovskaya, A.~Osipov, M.~Gerke, I.~Chestnov, S.~Arakelian,
  A.~S. Shalin, A.~B. Evlyukhin and A.~V. Kavokin , \emph{Scientific Reports},
  2019, \textbf{9}, 338\relax
\mciteBstWouldAddEndPuncttrue
\mciteSetBstMidEndSepPunct{\mcitedefaultmidpunct}
{\mcitedefaultendpunct}{\mcitedefaultseppunct}\relax
\EndOfBibitem
\bibitem[Naiki \emph{et~al.}(2011)Naiki, Masuo, Machida, and
  Itaya]{SinglePhoton_emission_JPCC_2011}
H.~Naiki, S.~Masuo, S.~Machida and A.~Itaya, \emph{The Journal of Physical
  Chemistry C}, 2011, \textbf{115}, 23299--23304\relax
\mciteBstWouldAddEndPuncttrue
\mciteSetBstMidEndSepPunct{\mcitedefaultmidpunct}
{\mcitedefaultendpunct}{\mcitedefaultseppunct}\relax
\EndOfBibitem
\bibitem[Focsan \emph{et~al.}(2014)Focsan, Gabudean, Vulpoi, and
  Astilean]{CdSe_Au_nanorod_Focsan_2014}
M.~Focsan, A.~M. Gabudean, A.~Vulpoi and S.~Astilean, \emph{The Journal of
  Physical Chemistry C}, 2014, \textbf{118}, 25190--25199\relax
\mciteBstWouldAddEndPuncttrue
\mciteSetBstMidEndSepPunct{\mcitedefaultmidpunct}
{\mcitedefaultendpunct}{\mcitedefaultseppunct}\relax
\EndOfBibitem
\bibitem[Dey \emph{et~al.}(2015)Dey, Zhou, Tian, Jenkins, Chen, Zou, and
  Zhao]{Multiexciton_emission_Zhao_2015}
S.~Dey, Y.~Zhou, X.~Tian, J.~A. Jenkins, O.~Chen, S.~Zou and J.~Zhao,
  \emph{Nanoscale}, 2015, \textbf{7}, 6851--6858\relax
\mciteBstWouldAddEndPuncttrue
\mciteSetBstMidEndSepPunct{\mcitedefaultmidpunct}
{\mcitedefaultendpunct}{\mcitedefaultseppunct}\relax
\EndOfBibitem
\end{mcitethebibliography}
\end{document}